\documentclass[aps,prd,twocolumn,groupedaddress,showpacs,nofootinbib,amssymb]{revtex4}
\usepackage[T1]{fontenc}
\usepackage[latin1]{inputenc}
\usepackage{graphicx}
\usepackage[english]{babel}
\usepackage{amsmath}
\usepackage{amssymb}
\usepackage{amsfonts}
\usepackage{colordvi}
\usepackage{color}
\begin{document}
\def\pp{{\, \mid \hskip -1.5mm =}}
\def\cL{{\cal L}}
\def\beq{\begin{equation}}
\def\eneq{\end{equation}}
\def\bea{\begin{eqnarray}}
\def\enea{\end{eqnarray}}
\def\tr{{\rm tr}\, }
\def\nn{\nonumber \\}
\def\e{{\rm e}}

\title{Noether Symmetry Approach for Dirac-Born-Infeld Cosmology}

\author{Salvatore Capozziello$^{1,2,3}$, Mariafelicia De Laurentis$^{4}$\footnote{e-mail address: mfdelaurentis@tspu.edu.ru}, Ratbay Myrzakulov$^{5}$,
%Daniele Vernieri$^{6}$
}

\affiliation{$^{1,2}$Dipartimento di Fisica, Università
di Napoli {}``Federico II'', Compl. Univ. di
Monte S. Angelo, Edificio G, Via Cinthia, I-80126, Napoli, Italy\\
INFN Sezione  di Napoli, Compl. Univ. di
Monte S. Angelo, Edificio G, Via Cinthia, I-80126, Napoli, Italy.}
\affiliation{$^3$Gran Sasso Science Institute (INFN), Via F. Crispi 7, I-67100, L' Aquila, Italy.}
\affiliation{$^{4}$Tomsk State Pedagogical University, 634061 Tomsk and National Research Tomsk State University, 634050 Tomsk, Russia}
\affiliation{$^5$ Eurasian International Center for Theoretical Physics and Department of General
Theoretical Physics, Eurasian National University, Astana 010008, Kazakhstan.}
%\affiliation{$^6$Centre National de la Recherche Scientifique (CNRS), IAP - Physique théorique: gravitation et cosmologie (GReCO)}
\date{\today}

\begin{abstract}
We consider the Noether Symmetry Approach  for a  cosmological model derived  from  a tachyon scalar field $T$ with a Dirac-Born-Infeld  Lagrangian and a potential $V(T)$. Furthermore,  we assume   a  coupled  canonical scalar field $\phi$ with an arbitrary interaction potential $B(T,\phi)$. Exact solutions are derived consistent with the  accelerated behavior of cosmic fluid.  
\end{abstract}
\maketitle
%%%%%%%%%%%%%
\section{introduction}
\label{1}
%%%%%%%%%
Scalar fields are introduced in cosmology in order to generate  dark energy dynamics and overcome shortcomings coming from the standard $\Lambda$CDM model.   Wide classes of theories involving  scalar fields have been considered in literature by  introducing different kinetic terms both canonical and  tachyonic and several interesting results come out in view to address the big puzzle of cosmic acceleration. Such scalar fields, in general, are obtained from particle physics and they can give rise to  quintessence or  phantom behavior in connection to the fact that they can be canonical or tachyonic \cite{7,8,9,10,11,12,12a1}.  In particular, there is much interest  in the tachyon cosmology since the presence of   tachyon fields is well motivated by string theory \cite{Chattopadhyay,9a,10a}. 
There are several work on the interaction between dark energy (tachyon or phantom) and dark matter, where  different kinds of interaction terms are phenomenologically introduced \cite{19a,20a,20b,21a,21b,21c}.
Specifically, tachyon fields come from the D-brane action (Dirac-Born-Infeld type Lagrangian) in string theory  and they  represent the lowest energy state in  unstable D-brane systems \cite{Gara,Gara1, Gara2, Becker, Zwiebach, Green}.  Features as   classical dust can be produced from an unstable D-brane through  pressureless gas with finite energy density \cite{11a,12a}. Some cosmological effects of a tachyon rolling down to its ground state have been widely discussed also using tachyon matter associated with unstable D-branes showing  an interesting equation of state as discussed in \cite{13a}. Furthermore,  tachyonic matter could provide an explanation for inflation at  early epochs  contributing to some new form of cosmological dark matter at late epochs \cite{14a}. Inflation derived from tachyonic fields has  been discussed in Ref.  \cite{15a,16a}. Furthermore, Dirac-Born-Infeld type action containing invariant curvature function, like $f(R, R^2, ...)$ gravity functions are considered in Refs. \cite{BDOlmo,davo,PhysRepnostro,OdintsovPR,faraoni}. 

Despite of these interesting results,  dynamical equations  deriving from this type of Lagrangians  are non-linear and general analytic solutions do not exist. Because of this fact,  the Noether Symmetry Approach could be a suitable method to seek for physically motivated solutions \cite{cimento,defelice,lambiase,lambiase1}.  
The existence of Noether symmetries leads to a specific form of coupling
function and the scalar field potential. Furthermore,
 symmetries lead to the existence of conserved quantities that allows to reduce dynamics thanks to the
presence of cyclic variables \cite{fGBnoether,greci,NS,fRTnoether}. The technique is used to obtain cosmological models in alternative gravity \cite{sharif,camci,sharif2,sharif3}. Black holes solutions have also been obtained by Noether symmetries \cite{noether-arturo,axially}.

This paper is organized as follow. 
In Sec. \ref{2} we introduce the Dirac-Born-Infeld type Lagrangian and derive  the cosmological  equations of motion. 
The Noether Symmetry Approach is described in  Sec. \ref{3}, and, in Sec. \ref{4},  it is   applied   to the Dirac-Born-Infeld  Lagrangian. Some exact solutions are found.  Conclusions are drawn in Sec. \ref{5}. 
%%%%%%%%%%%%%%%%%%%%%%%%%%%%%%%%%%%%%%
\section{Dirac-Born-Infeld type Lagrangian and equations of motion}
\label{2}
%%%%%%%%%%%%%%%%%%%%%%%%%%%%%%%%%%%%%%
Let us consider  a system of two interacting scalar fields, a tachyonic field $T$ and a canonical scalar field $\phi$, described by a Dirac-Born-Infeld type Lagrangian which couples the two fields through a potential $B(T,\phi)$. On the other hand, $V(T)$ is the tachyonic potential. 
The Lagrangian of the system can be written as \cite{Filo,filobello}
\beq
\mathcal{L}=-V(T)\sqrt{1-\partial_\mu T\partial^\mu T}+\frac{1}{2}\partial_\mu\phi\partial^\mu\phi-B(T,\phi).
\eneq

For a spatially flat Friedman-Robertson-Walker universe, we  obtain the following point-like Lagrangian
\beq
\mathcal{L}=-a^3V(T)\sqrt{1-\dot{T}^2}+\frac{1}{2}\dot{\phi}^2a^3-a^3B(T,\phi),     \label{4.4}
\eneq
where $a$ is the cosmic scale factor of the universe.
The Euler-Lagrange equations  give  dynamics for $\phi$ and $T$, that is 
\beq
\ddot{\phi}+3H\dot{\phi}+B_\phi(T,\phi)=0,                       \label{4.5}
\eneq
\beq
\frac{\ddot{T}}{\dot{T}^2-1}+3H\dot{T}+\frac{V_T(T)}{V(T)}+B_T(T,\phi)\frac{\sqrt{1-\dot{T}^2}}{V(T)}=0,      \label{4.6}
\eneq
where ${\displaystyle H=\frac{\dot{a}}{a}}$ is the Hubble parameter.  The subscripts `$T$' and `$\phi$' indicate  derivatives of the potentials $B(T,\phi)$ and $V(T)$ with respect to the two  fields. Dot indicates  derivatives with respect to the cosmic  time.  The equation for the energy is 
\begin{eqnarray}
E_{\cal L}=\frac{V(T){\dot T}^2}{\sqrt{1-\dot{T}^2}}+{\dot \phi}^2=0
\end{eqnarray}

%%%%%%%%%%%
\section{Noether symmetry approach}
\label{3}
%%%%%%%%%%%
Solutions for the dynamics given by the Lagrangian  (\ref{4.4}) can be achieved by selecting cyclic variables related to some Noether symmetry \cite{cimento, defelice,lambiase,lambiase1}. In principle, this approach allows to select models compatible with the symmetry so that  the conserved quantities could be considered  as a sort of Noether charges. Before going into the details of our model,  let us introduce  the Noether Symmetry Approach.

Let $\mathcal{L}(q^i,\dot{q}^i)$ be a canonical, independent of  time and non degenerate point-like Lagrangian where 
\beq
\frac{\partial\mathcal{L}}{\partial\lambda}=0, \hspace{1.5cm}  det H_{ij}\equiv \left\|\frac{\partial^2\mathcal{L}}{\partial \dot{q}^i\partial \dot{q}^j}\right\|\neq0,
\eneq
with $H_{ij}$ the Hessian matrix related to the Lagrangian $\mathcal{L}$ and  dot the derivative with respect to an affine parameter $\lambda$, which, in our case, corresponds to the cosmic time $\textit{t}$. In analytic mechanics, $\mathcal{L}$ is of the form
\beq
\mathcal{L}=K(\textbf{q},\dot{\textbf{q}})-V(\textbf{q}),                   \label{4.39}
\eneq
where $K$ and $V$ are the {\it kinetic energy} and {\it potential energy} respectively. We stress that ${\cal L}$ could assume more complicated expressions than  Eq. (\ref{4.39})  as in \cite{fGBnoether,greci,NS}. $K$ is a positive definite quadratic form in $\dot{\textbf{q}}$. The energy function associated with $\mathcal{L}$ is:
\beq
E_\mathcal{L}\equiv\frac{\partial\mathcal{L}}{\partial \dot{q}^i}\dot{q}^i-\mathcal{L},   \label{en}
\eneq
which , when the Lagrangian is in the form (\ref{4.39}), reduces to the total energy $K + V$. In any case, $E_\mathcal{L}$ is a constant of motion. Since our cosmological problem has a finite number of degrees of freedom, we are going to consider only point transformations.
Any invertible transformation of the `generalized positions' $Q^i=Q^i(\textbf{q})$ induces a transformation of the `gene-ralized velocities' such that:
\beq
\dot{Q}^i(\textbf{q})=\frac{\partial Q^i}{\partial q^j}\dot{q}^j;   \label{4.23}
\eneq
the matrix $\mathcal{J}=\left\|\partial Q^i/\partial q^j\right\|$ is the Jacobian of the transformation on the positions, and it is assumed to be non-zero. The Jacobian $\tilde{\mathcal{J}}$ of the induced transformation is easily derived and $\mathcal{J}\neq0\rightarrow\tilde{\mathcal{J}}\neq0$.
In general, this condition is not satisfied over the whole space but only in the neighborhood of a point. It is a local transformation.
A point transformation $Q^i=Q^i(\textbf{q})$ can depend on a (or more than one) parameter. As starting point, we can assume that a point transformation depends on a parameter $\epsilon$, so that $Q^i=Q^i(\textbf{q},\epsilon)$, and that it gives rise to a one-parameter Lie group. For infinitesimal values of $\epsilon$, the transformation is then generated by a vector field: for instance, $\partial/\partial x$ is a translation along the $x$ axis while $x(\partial/\partial y)-y(\partial/\partial x)$ is a rotation around the $z$ axis and so on.
In general, an infinitesimal point transformation is re-presented by a generic vector field on $Q$
\beq
\textbf{X}=\alpha^i(\textbf{q})\frac{\partial}{\partial q^i}.
\eneq
The induced transformation (\ref{4.23}) is then represented by
\beq
\textbf{X}^c=\alpha^i\frac{\partial}{\partial q^i}+\left(\frac{d}{d\lambda}\alpha^i(\textbf{q})\right)\frac{\partial}{\partial \dot{q}^j}.   \label{4.24}
\eneq
$\textbf{X}^c$ is called the \textit{complete lift} of \textbf{X}.
A function $F(\textbf{q}, \dot{\textbf{q}})$ is invariant under the transformation \textbf{X} if
\beq
L_XF\equiv\alpha^i(\textbf{q})\frac{\partial F}{\partial q^i}+\left(\frac{d}{d\lambda}\alpha^i(\textbf{q})\right)\frac{\partial}{\partial \dot{q}^j}F=0,
\eneq
where $L_XF$ is the Lie derivative of \textit{F}.
Specifically, if $L_X\mathcal{L}=0$, \textbf{X} is a symmetry for the dynamics derived by $\mathcal{L}$.
At this point let us consider a Lagrangian $\mathcal{L}$ and its Euler-Lagrange equations
\beq
\frac{d}{d\lambda}\frac{\partial\mathcal{L}}{\partial \dot{q}^j}-\frac{\partial\mathcal{L}}{\partial q^j}=0.                \label{4.25}
\eneq
Let us consider also the vector field (\ref{4.24}) and, contracting (\ref{4.25}) with $\alpha^i$s, it gives
\beq
\alpha^j\left(\frac{d}{d\lambda}\frac{\partial\mathcal{L}}{\partial \dot{q}^j}-\frac{\partial\mathcal{L}}{\partial q^j}\right)=0.     \label{4.26}      
\eneq
Since we have
\beq
\alpha^j\frac{d}{d\lambda}\frac{\partial\mathcal{L}}{\partial \dot{q}^j}=\frac{d}{d\lambda}\left(\alpha^j\frac{\partial\mathcal{L}}{\partial \dot{q}^j}\right)-\left(\frac{d\alpha^j}{d\lambda}\right)\frac{\partial\mathcal{L}}{\partial \dot{q}^j},
\eneq
from Eq. (\ref{4.26}),  it results that
\beq
\frac{d}{d\lambda}\left(\alpha^j\frac{\partial\mathcal{L}}{\partial \dot{q}^j}\right)=L_X\mathcal{L}.
\eneq
The immediate consequence is the {\it Noether theorem} which states:
if $L_X\mathcal{L}=0$, then the function
\beq
\Sigma_0=\alpha^k\frac{\partial\mathcal{L}}{\partial \dot{q}^k},                                                \label{4.27}
\eneq
is a constant of motion.
Some mathematical comments are necessary now. Eq. (\ref{4.27}) can be expressed independently of coordinates as a contraction of \textbf{X} by a Cartan $1$-form
\beq
\theta_\mathcal{L}\equiv\frac{\partial\mathcal{L}}{\partial \dot{q}^j}dq^j.
\eneq
For a generic vector field $\textbf{Y}=y^i\partial/\partial x^i$, and 1-form \\$\beta=\beta_idx^i$, we have, by definition, $i_Y\beta=y^i\beta_i$. Thus Eq. (\ref{4.27}) can be expressed as:
\beq
i_X\theta_\mathcal{L}=\Sigma_0.
\eneq
Through a point transformation, the vector field \textbf{X} becomes:
\beq
\tilde{\textbf{X}}=\left(i_XdQ^k\right)\frac{\partial}{\partial Q^k}+\left(\frac{d}{d\lambda}\left(i_XdQ^k\right)\right)\frac{\partial}{\partial \dot{Q}_k}.
\eneq
We see that $\tilde{\textbf{X}}$ is still the lift of a vector field defined on the space of configurations. If $\textbf{X}$ is a symmetry and we choose a point transformations such that
\beq
i_XdQ^1=1; \hspace{1.2cm}   i_XdQ^i=0 \hspace{1cm}  i\neq1,                           \label{4.38}
\eneq
we get
\beq
\tilde{\textbf{X}}=\frac{\partial}{\partial Q^1};  \hspace{1.2cm}   \frac{\partial \mathcal{L}}{\partial Q^1}=0.
\eneq
Thus $Q^1$ is a cyclic coordinate and the dynamics results reduced.
Furthermore, the change of coordinates given by (\ref{4.38}) is not unique and then a clever choice could be very important. In general, the solution of Eq. (\ref{4.38}) is not defined on the whole space. It is local in the sense explained above.
Considering the specific case which we are going to analyze,  the situation is the following. The Lagrangian is an application
\beq
\mathcal{L}:\mathcal{TQ}\rightarrow\mathbb{R},
\eneq
where $\mathbb{R}$ is the set of real numbers and the generator of symmetry is
\beq
\textbf{X}=\alpha\frac{\partial}{\partial T}+\beta\frac{\partial}{\partial \phi}+\gamma\frac{\partial}{\partial a}+\dot{\alpha}\frac{\partial}{\partial \dot{T}}+\dot{\beta}\frac{\partial}{\partial \dot{\phi}}+\dot{\gamma}\frac{\partial}{\partial \dot{a}}.           \label{4.28}
\eneq
As discussed above, a symmetry exists if the equation $L_X\mathcal{L}=0$ has solutions. Then there will be a constant of motion on shell, {\it i.e.} for the solutions of the Euler equations, as stated above Eq. (\ref{4.27}).
In other words, a symmetry exists if at least one of the functions $\alpha$, $\beta$ or $\gamma$ in Eq.(\ref{4.28}) is different from zero.
%%%%%%%%%%%%%
\section{Noether symmetry approach for Dirac-Born-Infeld type-Lagrangian}
\label{4}
\begin{figure}[h!]
\thicklines
\includegraphics[scale=0.40]{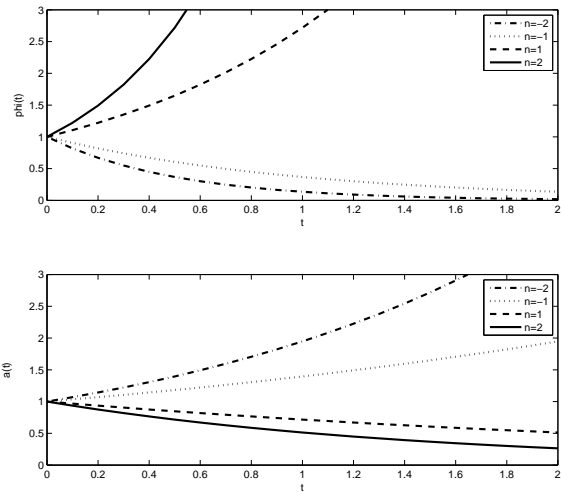}
\caption {In the top figure is plotted the behavior of $\phi(t)$ and  in the bottom figure the behavior of $a(t)$ for different values of $n$ respectively for the solution \eqref{10}. The solid lines represent the behavior for $n=-2$, the dashed  for $n=-1$, the dotter for $n=1$ and dashed-dot for $n=2$. }\label{fig1}
\end{figure}

\begin{figure}[h!]
\thicklines
\includegraphics[scale=0.48]{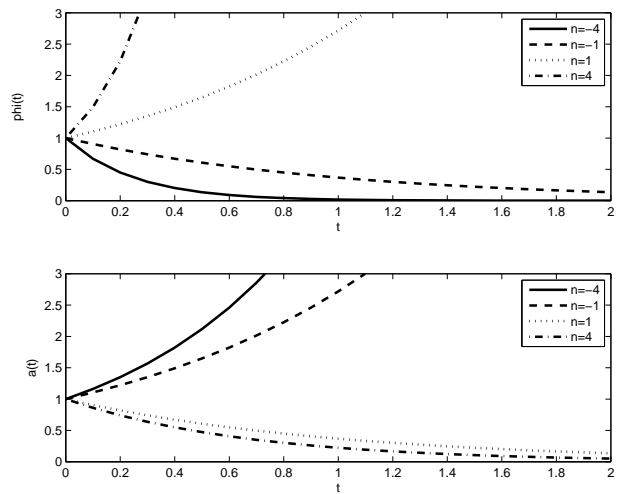}
\caption {Plots of $\phi(t)$ and $a(t)$ for the solution \eqref{11} for different values of $n$. The solid lines represent the behavior for $n=-4$, the dashed for $n=-1$, the dotter for $n=1$ and dashed-dot for $n=4$}\label{fig2}
\end{figure}

%%%%%%%%%
Let us apply now the method described in the previous section to the Lagrangian in Eq. (\ref{4.4}). 
From the statement $L_X\mathcal{L}=0$, we obtain:
\bea
&\alpha&\left[-a^3V_T(T)\sqrt{1-\dot{T}^2}-a^3B_T(T,\phi)\right]-a^3\beta B_\phi(T,\phi) \nonumber \\
&+&3a^2\gamma\left[-V(T)\sqrt{1-\dot{T}^2}+\frac{1}{2}\dot{\phi}^2-B(T,\phi)\right] \nonumber   \\
&+&\left(\dot{\phi}\partial_\phi\alpha+\dot{T}\partial_T\alpha +\dot{a}\partial_a\alpha \right)\left(\frac{a^3V(T)\dot{T}}{\sqrt{1-\dot{T}^2}}\right) \nonumber  \\
&+&\left(\dot{\phi}\partial_\phi\beta+\dot{T}\partial_T\beta +\dot{a}\partial_a\beta\right)\left(\dot{\phi}a^3\right)=0.
\enea
Setting to zero the coefficients of the terms $\dot{\phi}^2$, $\dot{T}^2$, $\dot{\phi}\dot{T}$, $\dot{a}\dot{\phi}$ and $\dot{a}\dot{T}$, we obtain respectively:
\beq
\frac{3}{2}\gamma a^2+a^3\frac{\partial\beta}{\partial\phi}=0,       \label{4.40}
\eneq
\beq
\frac{\partial\alpha}{\partial T}=0,                  \label{4.41}
\eneq
\beq
a^3\frac{\partial\beta}{\partial T}+\left(\frac{a^3V(T)\dot{T}}{\sqrt{1-\dot{T}^2}}\right)\frac{\partial\alpha}{\partial \phi}=0,    \label{4.42}
\eneq
\beq
\frac{\partial\beta}{\partial a}=0, \hspace{1cm} \frac{\partial\alpha}{\partial a}=0.     \label{4.43}
\eneq
Finally we have to satisfy the constraint 
\bea
&&a\alpha \left[V_T(T)\sqrt{1-\dot{T}^2}-B_T(T,\phi)\right]+a\beta B_\phi(T,\phi) +\nonumber \\      
&+&3\gamma \left[V(T)\sqrt{1-\dot{T}^2}+B(T,\phi)\right]=0.       \label{4.44}
\enea
Eqs. (\ref{4.40})-(\ref{4.43}) are consistent for constant values 
\beq
\alpha(a,\phi,T)=\alpha_0,
\eneq 
\beq\beta(a,\phi,T)=\beta_0,
\eneq
and finally 
\beq
\gamma(a,\phi,T)=0.
\eneq
The existence of non-zero quantities $\alpha$ and $\beta$ accounts for the Noether symmetry, provided that they satisfy the constraint (\ref{4.44}), which now becomes
\beq
\alpha_0\left[V_T(T)\sqrt{1-\dot{T}^2}+B_T(T,\phi)\right]+\beta_0 B_\phi(T,\phi)=0.   \label{constraint}
\eneq
It is evident that this constraint makes a direct connection between the tachyon potential $V(T)$ and the potential $B(T,\phi)$. 
Finally, we find the constant of motion, namely the Noether charge, as
\beq
\Sigma_0=\alpha_0\left(\frac{a^3V(T)\dot{T}}{\sqrt{1-\dot{T}^2}}\right)+\beta_0a^3\dot{\phi}.
\eneq
As we can see the constant of motion connects directly the dynamical evolution of the scalar field $\phi$ and  the tachyon field $T$ only through the tachyonic potential $V(T)$, and not  the coupled potential $B(T,\phi)$.
Considering the system of Eqs. (\ref{4.5}), (\ref{4.6}) and (\ref{constraint}), we can find some particular solutions making appropriate positions.
In fact supposing that the tachyon scalar field does not depend on time, being $T=T_0$ with $T_0$ a constant, the system of equations is solved for
\bea
\phi(t)&=&\phi_0e^{nt}, \nonumber \\
a(t)&=&a_0e^{-\frac{1}{3}nt}, \nonumber \\
B_\phi&=&0, \nonumber \\
B_T(T_0)&=-&V_T(T_0),    \label{10}
\enea
where $\phi_0$ and $n$ are arbitrary constants while $\displaystyle{a_0=\left(\frac{\Sigma_0}{\beta_0\phi_0 n}\right)^{\frac{1}{3}}}$.
A second  solution can be achieved considering the tachyon scalar field to be constant, being $T=T_0$, and $\Sigma_0=\beta_0=0$. In this case the system of equations is solved for
\bea
\phi(t)&=&\phi_0e^{nt}, \nonumber \\
a(t)&=&a_1e^{-\left(\frac{2+n^2}{3n}\right)t}, \nonumber \\
B_T&=&0, \nonumber \\
B(t)&=&B_0e^{2nt}, \nonumber \\
V&=&V_0, \label{11}\enea
where $\phi_0$, $n$, $a_1$, $B_0$ and $V_0$ are arbitrary constants.
Clearly, the expansion explicitly depends on $n$. In Figs. \ref{fig1} and \ref{fig2} are showed the trends of the potential and of the scale factor for different values of $n$.  We are adopting a dimensionless time parameter where $t=1$ corresponds to a Hubble time $H_0^{-1}$. 

%%%%%%%%%%%%%
\section{Conclusions}
\label{5}
%%%%%%%%%%%%%%%%%%%%%
In this paper, we have obtained the dynamics equations and investigated the conditions for the existence of a Noether symmetry in a Dirac-Born-Infeld type Lagrangian with a tachyonic potential $V(T)$ coupled to a canonical scalar field $\phi$ through an arbitrary interaction potential $B(T,\phi)$.
We have shown that a Noether symmetry  exists and is related to a constant of motion. 
Then, we have considered two particular cases where the tachyon scalar field $T$ is a  constant and  solved the system of equations. 
In the first case, it is possible to find particular solutions for the system of Eqs. (\ref{4.5}), (\ref{4.6}) and (\ref{constraint}), obtaining the  explicit behavior of the scale factor $a(t)$ and of the canonical scalar field $\phi(t)$. As it is clear  in the solutions (\ref{10}),  both of them  have exponential behavior, while the coupling potential $B$ is  independent of $\phi$. From the condition existing between the derivatives of $B$ and $V$ with respect $T$, calculated in $T=T_0$, it is possible to obtain a constant value for the tachyonic field. 
In the second case we find the solutions (\ref{11}), where $a(t)$, $\phi(t)$ and $B(t)$ are exponential functions.  In a forthcoming paper, we will develop these considerations to more general potentials $V(T)$ and $B(T,\phi)$.

%%%%%%%%%

\end{document}